\documentclass[aps,prb,superscriptaddress,twocolumn]{revtex4-2}

\usepackage{setspace}
\usepackage{lmodern}

\usepackage[latin9]{inputenc}
\usepackage{amsmath}
\usepackage{amssymb}
\usepackage{graphicx}
\usepackage{epstopdf}
\usepackage{color}
\usepackage{enumerate}
\usepackage{ulem}
\usepackage{bm} 
\usepackage{physics}

\usepackage{array}
\usepackage[labelfont=bf]{caption}
\usepackage{multirow}
\usepackage{floatpag}
\usepackage[misc]{ifsym}
\usepackage{sistyle}

\newcommand{\vf}{v_\text{F}}

\begin{document}

\title{Absence of edge reconstruction for quantum Hall edge channels in graphene devices}

\author{Alexis Coissard}
\author{Adolfo G. Grushin}
\affiliation{Univ. Grenoble Alpes, CNRS, Grenoble INP, Institut N\'{e}el, 38000 Grenoble, France}
\author{C\'ecile Repellin}
\affiliation{Univ. Grenoble Alpes, CNRS, LPMMC, 38000 Grenoble, France}
\author{Louis Veyrat}
\affiliation{Univ. Grenoble Alpes, CNRS, Grenoble INP, Institut N\'{e}el, 38000 Grenoble, France}
\author{Kenji Watanabe}
\affiliation{Research Center for Functional Materials, National Institute for Materials Science, 1-1 Namiki, Tsukuba 305-0044, Japan}
\author{Takashi Taniguchi}
\affiliation{International Center for Materials Nanoarchitectonics, National Institute for Materials Science,  1-1 Namiki, Tsukuba 305-0044, Japan}
\author{Fr\'{e}d\'{e}ric Gay}
\author{Herv\'e Courtois}
\author{Hermann Sellier}
\author{Benjamin Sac\'{e}p\'{e}}
\email{benjamin.sacepe@neel.cnrs.fr}
\altaffiliation[Present address: ]{Google Quantum AI, Mountain View, CA, USA}
\affiliation{Univ. Grenoble Alpes, CNRS, Grenoble INP, Institut N\'{e}el, 38000 Grenoble, France}

\begin{abstract}
\bf{Electronic edge states in topological insulators have become a major paradigm in physics. The oldest and primary example is that of quantum Hall (QH) edge channels that propagate along the periphery of two-dimensional electron gases (2DEGs) under perpendicular magnetic field. Yet, despite 40 years of intensive studies using a variety of transport and scanning probe techniques, imaging the real-space structure of QH edge channels has proven difficult, mainly due to the buried nature of most 2DEGs in semiconductors. Here, we show that QH edge states in graphene are confined to a few magnetic lengths at the crystal edges by performing scanning tunneling spectroscopy up to the edge of a graphene flake on hexagonal boron nitride. These findings indicate that QH edge states are defined by boundary conditions of vanishing electronic wavefunctions at the crystal edges, resulting in ideal one-dimensional chiral channels, free of electrostatic reconstruction. We further evidence a uniform charge carrier density at the edges, contrasting with conjectures on the existence of non-topological upstream modes. The absence of electrostatic reconstruction of quantum Hall edge states has profound implications for the universality of electron and heat transport experiments in graphene-based systems and other 2D crystalline materials.}
\end{abstract}

\maketitle
\section*{Introduction}

In 1982, two years after the discovery of the quantum Hall effect~\cite{Klitzing1980}, B. Halperin predicted the existence of edge states carrying the electron flow along sample periphery~\cite{Halperin82}. These edge states, which form unidirectional (chiral) ballistic conduction channels, have been pivotal in understanding most of the transport properties of the QH effect~\cite{Buttiker88,Beenakker11}. They have served as an extraordinarily versatile platform for a multitude of quantum coherent experiments~\cite{bauerle18}, culminating recently in the evidence of fractional statistics in the fractional QH effect~\cite{Bartolomei20} and the possibility of anyon braiding through interferometry~\cite{Nakamura20}.

\begin{figure*}[ht]
\center
\includegraphics[width=0.9\linewidth]{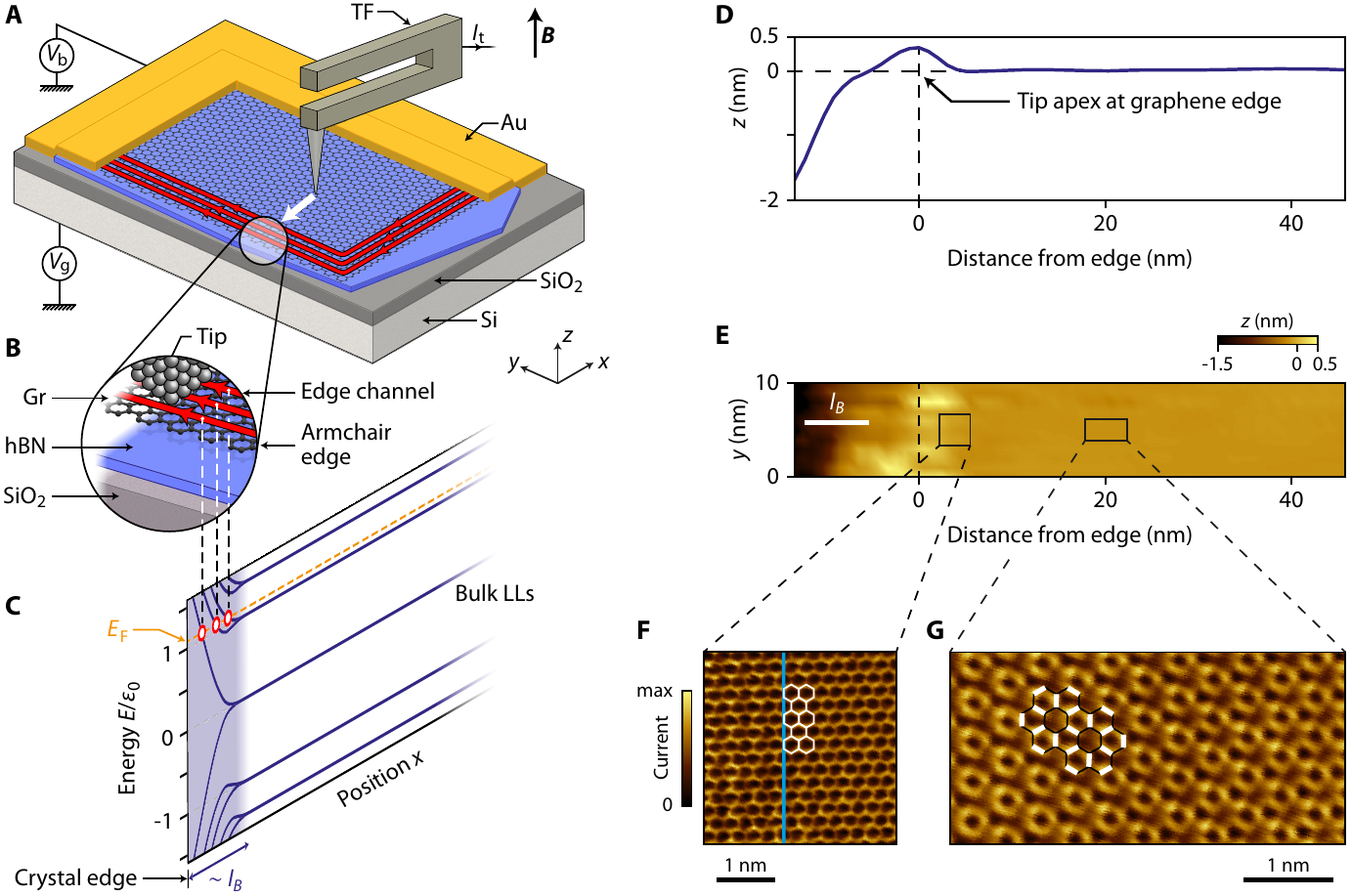}
\caption{\textbf{Tunneling spectroscopy of quantum Hall edge states.} \textbf{A, B}, Schematics of the experiment. A PtIr tip is glued at the extremity of one prong of a piezoelectric tuning fork to enable imaging both in STM (by regulating the tunneling current $I_\text{t}$) and in AFM (by regulating the frequency shift of the tuning fork). Graphene lies atop an insulating hBN flake and is contacted by a Cr/Pt/Au electrode to apply the sample bias $V_\text{b}$. A back-gate voltage $V_\text{g}$ applied to the Si/SiO$_2$ substrate enables to tune the Fermi level $E_F$ in graphene. Graphene edges are first located by AFM under perpendicular magnetic field, $B$. The tip is then moved from the graphene bulk to the edge in STM to perform tunneling spectroscopy of QH edge channels. \textbf{C}, Landau level spectrum~\cite{Abanin2006,Brey06,Abanin2007b} as a function of energy $E$ (normalized to the first cyclotron gap $\varepsilon_0$) and position. The Landau levels disperse at an armchair edge on the scale of the magnetic length $l_B$. Their intersect with the Fermi level defines the QH edge channels. \textbf{D, E}, Topographic image (E) and its $z$ profile averaged on the $y$ direction (D) of the graphene edge obtained in STM. We consider that the tip apex is located above the graphene edge at the maximum of the $z$ profile. \textbf{F}, Atomic resolution of the graphene honeycomb lattice measured in STM a few nanometers away from the edge. The vertical blue line indicates the crystal edge orientation deduced from (E). \textbf{G}, Kekul\'e-bond order imaged in charge-neutral graphene~\cite{Coissard22} at $V_\text{g}$=-5\:$V$ at a distance of $20\:$nm from the edge.}
\label{Fig1}
\end{figure*}

The existence of edge states was initially inferred as a consequence of the boundary conditions imposed by the physical edges on the electron wavefunctions~\cite{Halperin82}. The energy of the electron states that are condensed into Landau levels increases upon approaching the edge due to the hard-wall boundary conditions, opening new conduction channels --the QH edge channels-- spatially located at their intersection with the Fermi level~\cite{Halperin82} (see Fig.~\ref{Fig1}B-C). Inclusion of a smooth electrostatic confining potential, which is experimentally used to define edges in 2DEGs buried in semiconductor heterostructures, enriches the picture with the concept of edge reconstruction~\cite{Chklovskii92}. There, the Coulomb interaction energy dominates the confining potential, leading to a transformation of the edge states into a series of wide compressible channels separated by incompressible strips. In the opposite case of a sharp potential, the Coulomb interaction is not relevant and the single particle picture is valid.
Edge reconstruction mechanisms have further proven to be of paramount importance in the fractional QH regime where additional co- and/or counter-propagative or even neutral modes~\cite{Chamon94,Kane94,Khanna21} can emerge and complexify charge and heat transport~\cite{Venkatachalam12,Goldstein16,Bhattacharyya19}.

Non-reconstructed edge states can substantially clarify QH edge transport with virtually ideal one-dimensional edge states~\cite{ZiXiang11} and new regimes of intra- and inter-channel interactions. Contrary to semiconductor heterostructures, two-dimensional crystalline materials like graphene, for which physical edges are crystal edges, may be archetypical systems hosting such edge states.  For graphene and its massless, linear band structure, QH edge states without confining electrostatic potential are expected to be the exact eigenstates of the Dirac equations derived with vanishing boundary conditions at the armchair or zigzag edge~\cite{Abanin2006,Brey06,Abanin2007b}.  Akin to Halperin's original prediction~\cite{Halperin82}, these solutions for edges states are maximally confined to a few magnetic lengths $l_B=\sqrt{\hbar/eB}$ ($\hbar$ is the reduced Planck constant, $e$ the electron charge and $B$ the magnetic field) from the crystal edge,  leaving no room for edge reconstruction.

Here, we unveil the real-space structure of the quantum Hall edge states of graphene lying on an insulating hexagonal boron nitride flake (hBN) and evidence the absence of edge reconstruction, by performing scanning tunneling spectroscopy up to the graphene crystal edge, under strong perpendicular magnetic field.  We achieved this by overcoming the long-standing experimental challenge~\cite{McCormick99,Yacoby1999,Weis2011,Ito2011,Lai2011,Suddards2012,Weitz2000,Nazin10,Li2013,Pascher14,Kim2020} of approaching a scanning tunneling tip to the edge without crashing it on the insulating substrate that borders the graphene flake, by means of a prior localization of the graphene edge by atomic force microscopy (AFM). We purposely used a home-made hybrid scanning microscope~\cite{Coissard22} capable of operating alternatively in AFM and scanning tunneling microscopy (STM) mode, thanks to a PtIr STM tip glued onto a piezoelectric tuning fork acting as a force sensor~\cite{Giessibl2004,Senzier07} for AFM (see Fig. \ref{Fig1}A).  
Our sample schematized in Fig. \ref{Fig1}A and B consists of a graphene monolayer deposited on a hBN flake sitting on a Si/SiO$_2$ substrate that serves as a back-gate electrode (see Methods). The graphene flake is contacted by a Cr/Pt/Au tri-layer that allows to apply a voltage bias $V_\text{b}$ and collect a tunnel current $I_\text{t}$ via the STM tip. All experiments presented here are performed at a temperature of 4.2~K and a perpendicular magnetic field of 14 T.
\\

\section*{Results}

\textbf{Quantum Hall edge states spectroscopy}

Figure \ref{Fig1}E displays a STM topographic image taken in constant current mode to the graphene edge, initially coarsely located by AFM (see Fig. S1). The height profile of this image (Fig. \ref{Fig1}D) shows a large flat area, and a slight bump on the left part of the scan. This bump results from the tip-graphene interaction lifting up the graphene edge when the tip is right above it~\cite{Georgi2017}. This bump allows us to locate the edge of the graphene crystal  with an accuracy of a few nanometers (see SI). To the left of the bump, the tip dips towards the hBN substrate, on which a tip crash is avoided by a height limit of the STM controller. Atomic scale imaging of the honeycomb lattice shown in Fig.~\ref{Fig1}F gives insight into the graphene lattice termination. The edge orientation in Fig.~\ref{Fig1}E, which is reported in Fig.~\ref{Fig1}F with the blue line, indicates an armchair termination. 

\begin{figure*}[ht!]
\includegraphics[width=0.9\linewidth]{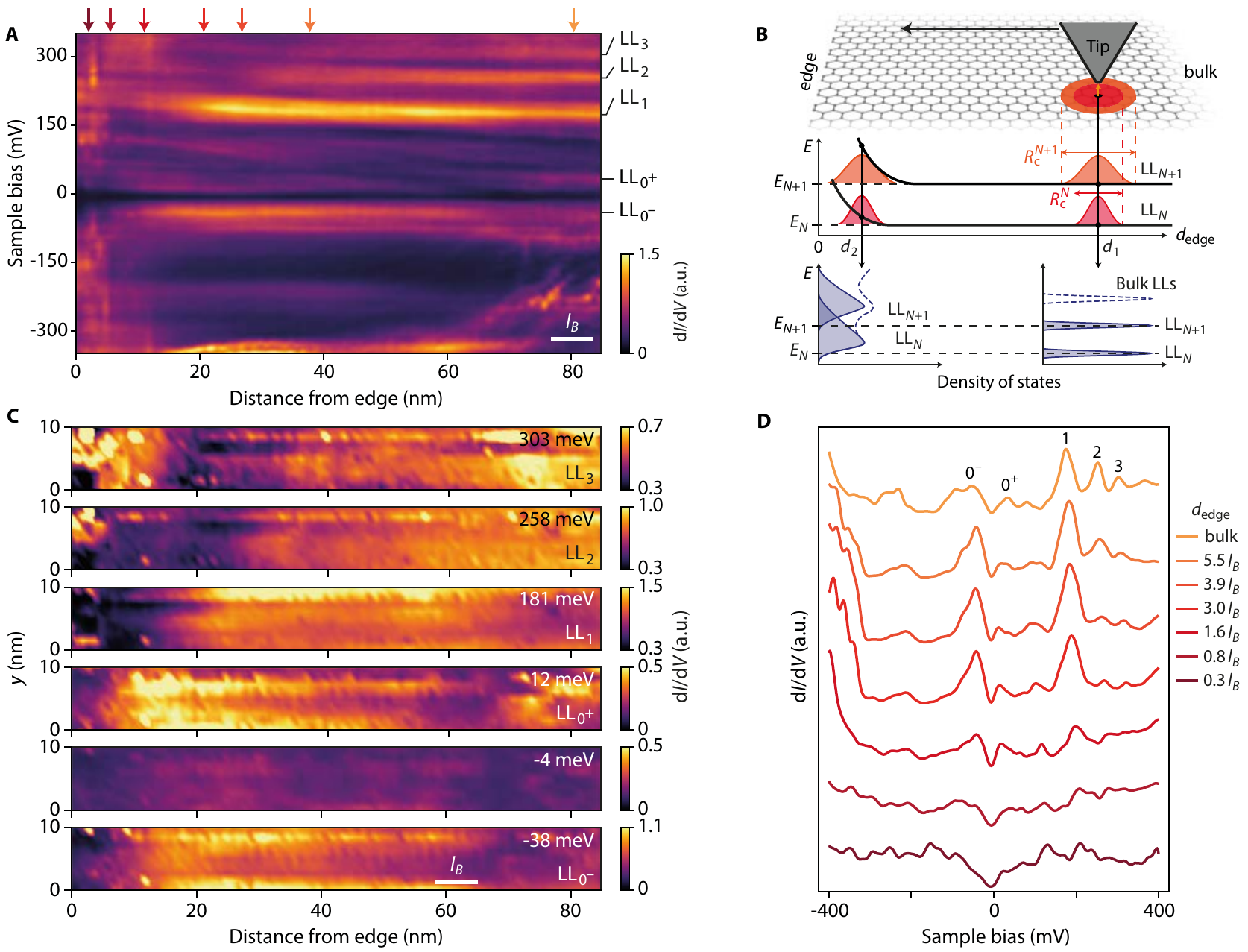}
\centering
\caption{\textbf{Sharp quantum Hall edge states.} \textbf{A}, Evolution of the tunneling conductance d$I_\text{t}$/d$V_\text{b}$ as a function of the distance from graphene edge measured at charge neutrality ($V_\text{g} = -5.4$~V). The half-filled zeroth Landau level is split into two sublevels LL$_{0^+}$ and LL$_{0^-}$ due to quantum Hall ferromagnetism~\cite{Coissard22}. \textbf{B}, Schematics of the tunneling into QH edge states. Due to the spatial extent $R_\text{c}^N=l_B(2|N|+1)^{1/2}$ of the LL$_N$ wave functions, the tunneling electrons probe at one point contributions from all states up to distances of about $R_\text{c}^N$ (red and orange gaussians in the middle panel for LL$_N$ and LL$_{N+1}$, respectively). The resulting density of states features sharp Landau level peaks in the bulk, i.e. at distance $d_1$ from the edge, and a smooth profile close to the edge, at a distance $d_2 \sim l_B$, due to the energy broadening of Landau levels \cite{Abanin2006}. Additionally, when approaching the edge, the tip starts to probe the edge states of the lower Landau levels, pushed at higher energies by the presence of the physical edge, and overlapping with the highly-degenerate bulk states. The resulting peaks in the density of states thus exhibit a spectral weight redistribution toward higher energies, which leads to a suppression of the Landau level peak height in the tunneling conductance (bottom panel, in solid blue each individual $N$ and $N+1$ Landau level peak, and in dashed blue the overall density of states). \textbf{C}, Spatial maps of the tunneling conductance d$I_\text{t}$/d$V_\text{b}$ at the energies of the Landau levels. \textbf{D}, Individual spectra taken from (A) at different distances from the edge indicated by the color-coded arrows in (A).}
\label{Fig2}
\end{figure*}

The central result of this work is shown in Figure \ref{Fig2}, which presents the evolution of the Landau levels upon approaching the immediate proximity of the graphene edge in the region shown in Fig.~\ref{Fig1}E, under a magnetic field of 14~T. We first study charge-neutral graphene by tuning the density with the back-gate voltage set at $V_\text{g} = -5.4$~V. Tunneling spectroscopy of Landau levels~\cite{Matsui05,Hashimoto08,Song10,Andrei2012} results in a series of peaks in the tunneling conductance $G(V_\text{b})=\text{d}I_\text{t}/\text{d}V_\text{b}$ that is proportional to the local density of states. We show in Fig. \ref{Fig2}A the tunneling conductance $G(d_\text{edge},V_\text{b})$ as a function of tip distance perpendicular to the graphene edge $d_\text{edge}$, and bias voltage $V_\text{b}$. Far from the edge, Landau levels are readily identified as bright conductance peaks that we label LL$_N$, where $N$ is the Landau level index.  These conductance peaks are conspicuously stable upon approaching the edge on the left of the figure. Within 40~nm from the edge, we observe a suppression of the Landau level peak heights (see individual spectra in Fig.~\ref{Fig2}D) starting at distances that depend on the Landau level (the higher the Landau index, the further from the edge). Figure ~\ref{Fig2}C shows spatial maps of the tunneling conductance at the voltage bias of the Landau level peaks. For each Landau level peak, darker areas corresponding to Landau level peak suppression appear further and further from the edge as the Landau level index increases.

These findings contrast with the expectation for a smooth confining potential at the edges, for which the Landau level spectrum would have continuously shifted in energy, following the confining potential as the edge is approached. Since the tunneling conductance probes states on the scale of the electron wavefunction, that is, the cyclotron radius $R_\text{c}^N=l_B(2|N|+1)^{1/2}$ for Landau level index $N$, the suppression of the Landau level peaks, here, reflects a spreading of the spectral weight to higher energy due to an abrupt edge state dispersion at the physical edge, on a very short scale of the order of the magnetic length (see Fig.~\ref{Fig2}B). This suppression of the tunneling density of states of the Landau levels, which has been observed on graphene on a conductive graphite substrate~\cite{Li2013}, is therefore direct evidence of QH edge states sharply confined at the edges. Ultimately, on the last few nanometers from the edge, the Landau level peaks disappear completely, and the redistribution of Landau level spectral weight yields a V-shape like tunneling density of states (see Fig.~\ref{Fig2}D).

\begin{figure*}[ht!]
\includegraphics[width=0.9\linewidth]{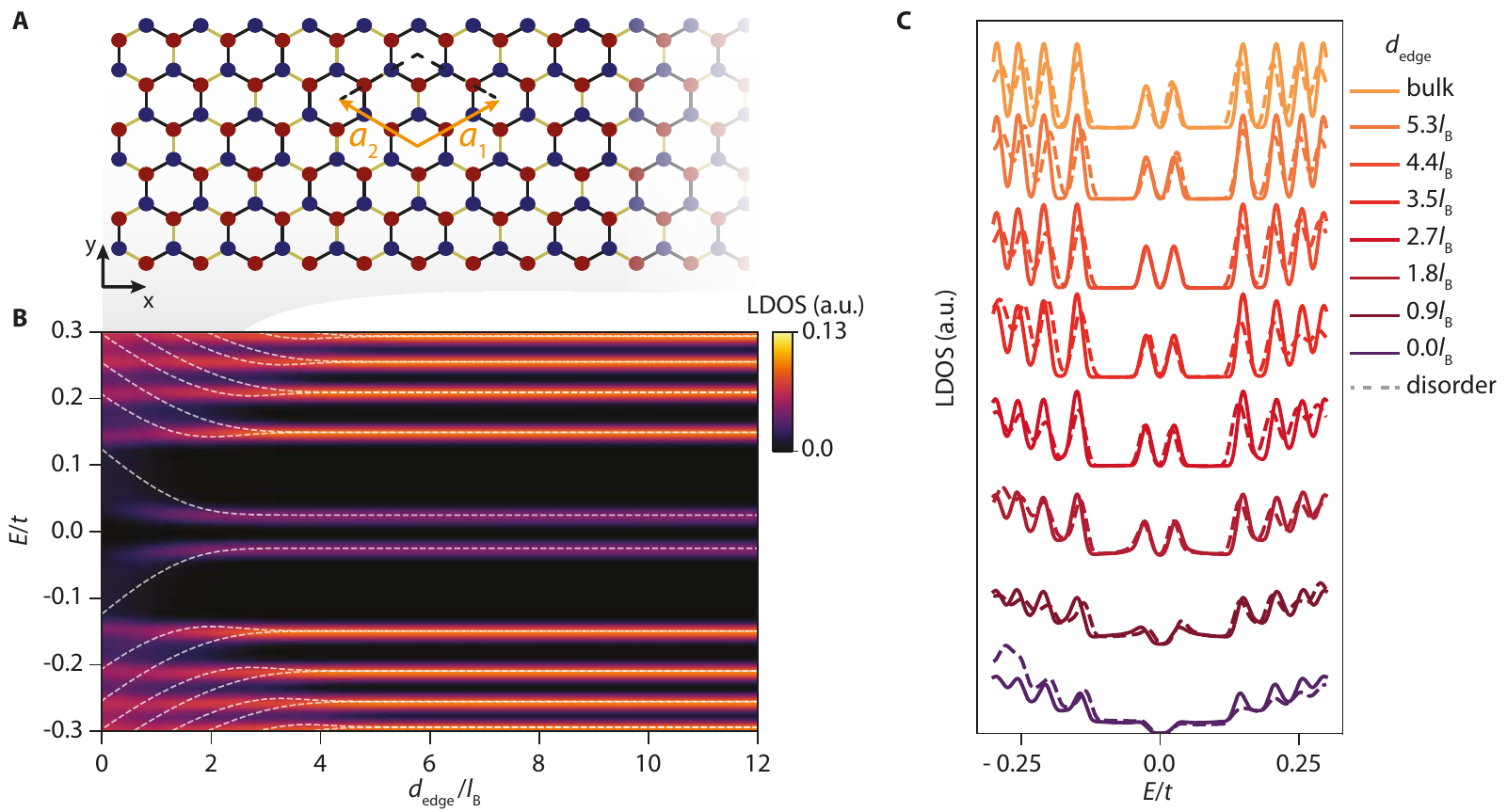}
\centering
\caption{\textbf{Theoretical tunneling density of states.} 
\textbf{A}, Schematic of the simulated edge-geometry.
We considered a Kekul\'{e}-bond order~\cite{Li2019,Liu2022,Coissard22} as broken-symmetry state~\cite{Goerbig22}, with lattice vectors that triple the unit cell compared to pristine graphene. 
\textbf{B}, Corresponding local density of states (LDOS) as a function of the distance from the armchair graphene edge, $d_{\mathrm{edge}}$, normalized by $l_B$, for charge-neutral graphene. The Kekul\'{e}-bond order splits the zeroth Landau level into two sublevels LL$_{0^+}$ and LL$_{0^-}$ with an energy gap chosen to match the experimentally measured value of $50$ m$e$V~\cite{Coissard22}. The white dashed lines are the numerically computed Landau levels of graphene nano-ribbon with armchair termination. Due to position momentum locking, the Landau Levels disperse as they approach the physical edge as sketched in Fig. \ref{Fig1}B. \textbf{C}, Individual spectra taken from (B) at different distances from the edge. Solid lines show cuts at different $d_{\mathrm{edge}}$, while dashed lines show spectral asymmetry emerging from a single disorder realization of an on-site disorder potential with strength $W/t= 0.3$ (see Methods).}
\label{Fig3}
\end{figure*}

In this measurement we have set the Fermi level at charge neutrality, that is, at Landau level filling factor $\nu=0$, which leads to a splitting of the zeroth Landau level (see split peaks labeled LL$_{0^+}$ and LL$_{0^-}$ in Fig.~\ref{Fig2}A) with the opening of an interaction-induced gap at $V_\text{b} = 0$~V (see Ref.~\cite{Coissard22}). This splitting signals the broken-symmetry state~\cite{Goerbig22} at charge neutrality with the Kekul\'{e}-bond order~\cite{Li2019,Liu2022,Coissard22}. Interestingly, we identified the Kekul\'{e}-bond order at 20~nm of the edge in Fig.~\ref{Fig1}G, indicating that this broken-symmetry state, which develops in the bulk, is robust even in the very proximity of the edge~\cite{Knothe15}.

To substantiate our finding we performed numerical simulations of the local density of states of a charge neutral graphene graphene ribbon with an armchair edge under perpendicular magnetic field (see. Fig. \ref{Fig3}A) \cite{Abanin2006,Brey06,Abanin2007b}.
We computed the Landau levels of the lattice Hamiltonian of nearest-neighbor hopping energy $t$. We assumed a Kekul\'{e}-bond order with a gap at half-filling of the zeroth Landau level of $50$ m$e$V, as measured experimentally~\cite{Coissard22}. The eigenstates for a ribbon with periodic boundary conditions along $\hat{y}$ are shown in Fig.~\ref{Fig3}B as white dashed lines.
The Landau level eigenstates disperse as their average $x$ position, locked to their $k_y$ momentum, approaches the physical edge of graphene~\cite{Pyatkovskiy14}, as schematized in Fig.~\ref{Fig1}B. Fig. \ref{Fig3}B shows the clean local density of states, which integrates the eigenstates weighted by the amplitude of the wavefunctions, as a function of the distance to the edge normalized by $l_B$, $d_{\mathrm{edge}}/l_B$, averaged over each unit cell (see Methods). The range of $d_{\mathrm{edge}}/l_B$ coincides with the range of displacement in Fig. \ref{Fig2}A, allowing direct comparison with the experimental data.  
The resulting Landau level peaks are suppressed at higher values of $d_{\mathrm{edge}}$ the higher their Landau level index, and on the same spatial scale as observed experimentally in Fig.~\ref{Fig2}A. 
This reduction of spectral weight is more visible in Fig.~\ref{Fig3}C where we plot spectra for different $d_{
\mathrm{edge}}$ (solid lines) including a single realization of on-site disorder (dashed lines).
The latter breaks the particle-hole symmetry of the spectrum, and thus may contribute to the asymmetries observed in the zeroth landau level peaks.
\\

\textbf{On the charge accumulation on the edges}

The question of charge carrier homogeneity is critical for graphene transport. A body of work has shown anomalous asymmetry in some transport properties supplemented by scanning probe investigations~\cite{Cui16,Marguerite19,Moreau21}, which points to a charge carrier accumulation at the graphene edges. Its origin may be either electrostatic stray field of the back-gate electrode~\cite{Silvestrov08} or chemical doping due to edge treatments (etching) or dangling bonds. In the QH effect, such an accumulation could open up additional counter-propagative edge channels and produce dissipation~\cite{Silvestrov08,Marguerite19,Moreau21}. 

\begin{figure*}[ht!]
\includegraphics[width=0.7\linewidth]{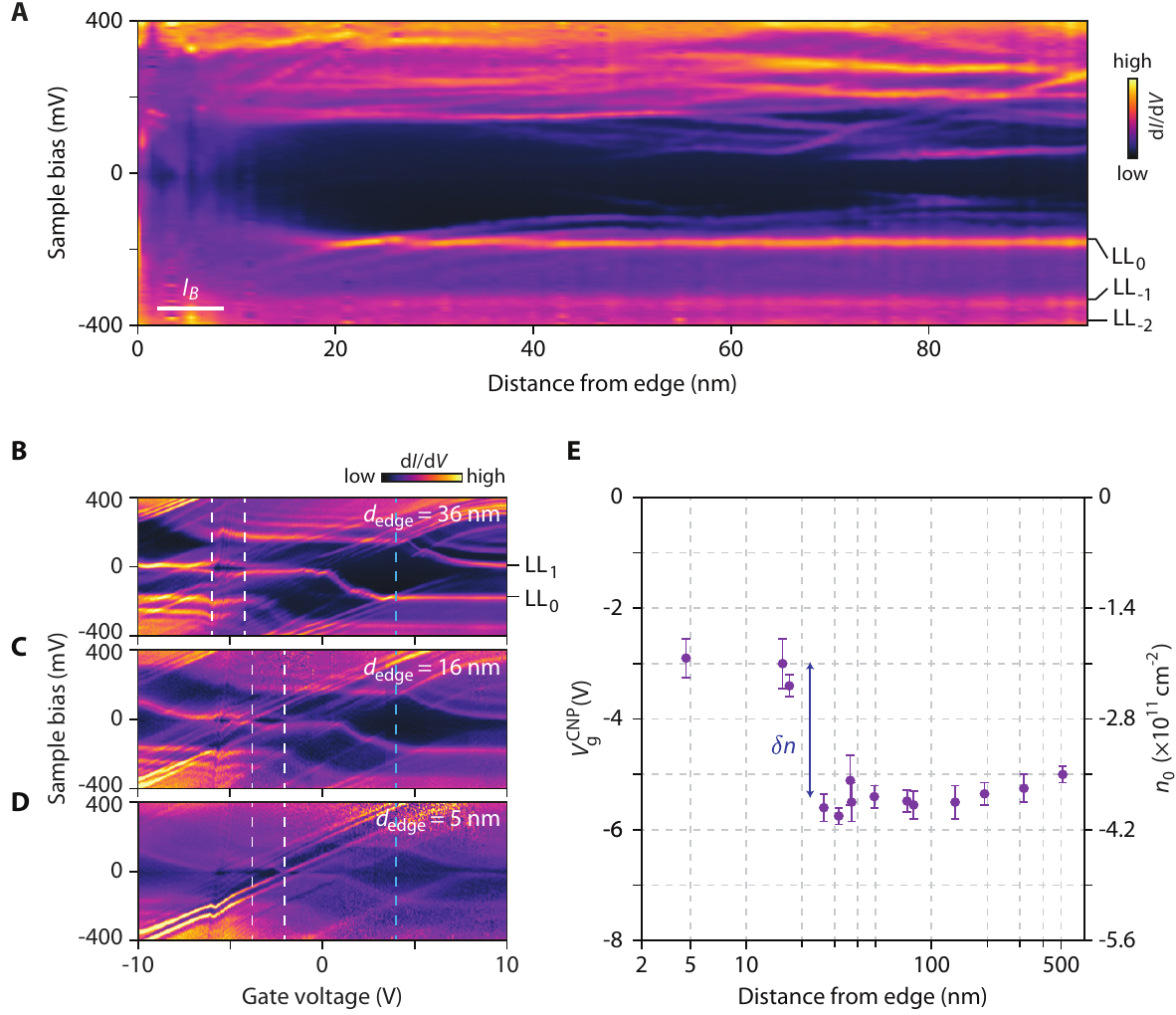}
\centering
\caption{\textbf{Charge density inhomogeneity on pristine edges.} \textbf{A}, Spatial evolution of the tunneling conductance up to the edge at filling factor $\nu=2$ for $V_\text{g} = 4$~V. \textbf{B-D}, Tunneling conductance gate maps as a function of sample bias $V_\text{b}$ and gate voltage $V_\text{g}$. At a distance $d_\text{edge}=36\:\text{nm}=5.3\:l_B$ from the edge, in (B), we observe the staircase pattern of Landau levels of the graphene bulk. At closer distances from the edge, the Landau level peaks in the staircase pattern starts to blur (C) and mostly vanishes in (D). In the three panels, the opening of the $\nu=0$ gap as a function of $V_\text{g}$ is indicated by white dashed lines, and the back-gate voltage of the charge-neutrality point $V_{\text{g}}^\text{CNP}$ is identified by the maximum of the gap. The blue dashed lines indicate the spectra at $V_{\text{g}} = 4 $ V, which coincide with the back-gate voltage of (A). The faint diagonal lines that translate into faint horizontal lines above the LL$_0$ peak in (A) result from residual charging effects in the tunneling process. Their downward dispersion near the edge in (A) is consistent with the charge carrier variation in (E). \textbf{E}, Evolution of $V_{\text{g}}^\text{CNP}$ and charge carrier density $n_0$ determined from tunneling conductance gate maps as a function of the distance from the edge $d_\text{edge}$.  The position of the charge carrier density shift coincides with the upward shift of the LL$_0$ in (A). Error bars correspond to the range of gate voltage where the $\nu=0$ gap opens in the gate maps.}
\label{Fig4}
\end{figure*} 

In tunneling experiments, a charge inhomogeneity on the edge would result in an energy shift of the Landau level spectrum as a whole due to a local change of the Landau level filling factor. Our measurements in Fig.~\ref{Fig2} provides a first insight on this issue with a remarkable stability of the Landau level peaks in energy that indicates that a possible charge accumulation is not large enough to depin the chemical potential from the zeroth Landau level~\cite{Coissard22}. In particular, it is lower than the value $\delta n = 6.8\times10^{11}$~cm$^{-2}$ required to fill the zeroth Landau level and reach $\nu=2$ at 14 T, which would produce a visible energy shift of the Landau level spectrum that we do not observe.

To enhance the sensitivity of the spectroscopy to possible charge inhomogeneities, we performed similar measurements at filling factor $\nu=2$ ($V_\text{g} = 4$~V), when the Fermi level is pinned by localized states in the cyclotron gap separating LL$_0$ from LL$_{1}$. There, due to the little density of localized states as compared to the highly degenerate Landau levels, a small variation of charge density would result in a substantial shift of the Landau levels in the tunneling spectra. Figure~\ref{Fig4} displays the spatial evolution of the tunneling conductance up to the edge, at $\nu=2$ and 14~T. As in Fig.~\ref{Fig2}, the Landau level peaks (LL$_0$, LL$_{-1}$ and LL$_{-2}$) stay at the same energy over the scan and vanish at about 20 nm from the edge, clearly indicating the absence of charge accumulation. We further performed systematic gate-tuned tunneling spectroscopy maps at various locations, from 500 nm to 5 nm from the edge (see SI). Figure~\ref{Fig4}B-D displays three of these maps taken close to the edge. We observe in Fig.~\ref{Fig4}B the usual staircase pattern of the Landau level peaks due to the successive pinning of the Fermi energy in the Landau levels~\cite{Luican2011,Chae2012,Coissard22}, which allows us to precisely identify the back-gate voltage of the charge neutrality point $V_\text{g}^{\text{CNP}}$. As shown in Fig.~\ref{Fig4}E that displays $V_\text{g}^{\text{CNP}}$ as a function of the distance from the edge, there is no charge accumulation from 500 nm to 20 nm to the edge, and only within 20 nm of the edge we measure a variation $\delta n =(-1.5\pm1.1)\times 10^{11}\:\text{cm}^{-2}$. 

Interestingly, such a charge density variation near the edge at 14~T yields a little variation $\delta \nu =0.4$ of local filling factor, which would have no consequence on the QH edge transport properties. Extrapolating at lower field, however, $\delta \nu =2$ would be reached at a magnetic field of 3~T, thus potentially affecting edge transport with additional modes.  Yet, the very small spatial scale of this charge accumulation cannot explain recent scanning probes experiments evidencing indirect, sometimes out-of-equilibrium responses within hundreds of nanometers from the edge~\cite{Marguerite19,Moreau21}. We conjecture that this charge accumulation in our particular case is related to the tip-graphene interaction when the tip reaches and lift up the graphene edge (see Supplementary Section III).
\\

\section*{Discussion}

The issue of charge accumulation on the edge and the ensuing emergence of upstream modes~\cite{Marguerite19,Moreau21} were put forth as an alternative interpretation~\cite{Aharon21} for the signature of helical edge transport in charge-neutral graphene~\cite{Jarillo2014,Veyrat2020}. Although we cannot exclude that the stray field of the back gate electrode may accumulate charges at high back-gate voltages, that is, away from charge neutrality point, and over a long distance~\cite{Silvestrov08}, our results show that this accumulation is absent at low back-gate voltage, thus invalidating the doubts raised~\cite{Aharon21} on the existence of the quantum Hall topological insulator phase in charge neutral-graphene~\cite{Jarillo2014,Veyrat2020}. Still, it may be interesting to revisit non-local transport in non-linear regime~\cite{Aharon21} in view of the exact spatial structure of the QH edge states in graphene.

Regarding edge reconstruction, a wealth of fractional and integer quantum Hall states exhibit complex sequences of reconstructed edge channels, including additional integer and/or fractional as well as neutral modes~\cite{Chamon94,Kane94,Khanna21}. Whereas the smooth electrostatic potential in GaAs and other semiconductors reconstructs edge states into wide compressible stripes of the order of $\sim 100$ nm (see Ref.~\cite{Pascher14}), the graphene QH edge states confined on a very short length scale, at few magnetic lengths on the physical edge, pose new constraints and limits for such a reconstruction, opening the investigation of universal transport and thermal properties~\cite{ZiXiang11}. Moreover, in such a strongly confined configuration, an enhancement of inter-edge-states interactions can be expected, which makes the picture of independent chiral channels irrelevant in this case, thus impacting charge and heat equilibration~\cite{Srivastav21,Kumar22,LeBreton22}. This should impact QH interferometry~\cite{Feldman21} in graphene systems~\cite{Deprez21,Ronen21} and other coherent experiments~\cite{bauerle18}, for which the independence, exact positions and nature of edge modes are crucial parameters to address anyon physics as well as other interaction-driven phenomena, such as charging effects~\cite{Halperin11}, spin-charge separation~\cite{Fujisawa22} or electron pairing~\cite{Choi2015}. 
\\

\vspace{1em}
\textit{Note}: A very recent work (https://arxiv.org/abs/2210.01831) reports a complementary tunneling spectroscopy study of electrostatically-defined QH edge states at a $pn$ junction.

\section*{methods}

\subsection*{Sample fabrication}

The graphene/hBN heterostructure was assembled from exfoliated flakes with the van der Waals pick-up technique using a polypropylene carbonate (PPC) polymer~\cite{Wang2013}. The stack with graphene on top of the hBN flake was deposited using the method described in Ref.~\cite{Nadj-Perge2019} on a highly p-doped Si substrate with a $285$~nm thick SiO$_2$ layer. Electron-beam lithography using a PMMA resist was used to pattern a guiding markerfield on the whole $5\times 5$ mm$^2$ substrate to drive the STM tip toward the device and to locate the graphene edge. Cr/Pt/Au electrodes contacting the graphene flake were also patterned by electron-beam lithography and metalized by e-gun evaporation. The sample was thermally annealed at $350\,^{\circ}$C in vacuum under an halogen lamp to remove resist residues and clean graphene, before being mounted into the STM where it was heated \textit{in situ} during the cooling to $4.2\;$K.

\subsection*{Measurements}

Experiments were performed with a home-made hybrid scanning tunneling microscope (STM) and atomic force microscope (AFM) operating at a temperature of $4.2\;$K in magnetic fields up to $14\;$T. The sensor consists of a hand-cut PtIr tip glued on the free prong of a tuning fork, the other prong being glued on a Macor substrate. Once mounted inside the STM, the tip is roughly aligned over the sample at room temperature. The AFM mode was used first for coarse navigation at $4.2\;$K on the sample surface to align the tip onto graphene and then for locating coarsely the graphene edge, see SI. The STM imaging in constant-height mode of the edge, done subsequently, yields a fine identification. Scanning tunneling spectroscopy (STS) was performed using a lock-in amplifier technique with a modulation frequency of $263\;$Hz and rms modulation voltage between $1-5\;$mV depending on the spectral range of interest. Current Imaging Tunneling Spectroscopy (CITS) measurements were acquired by starting far from the edge, with a grid whose slow $x$-axis is perpendicular to the edge direction (as imaged by STM) and the $y$-axis is parallel to the edge with a size of a few tens of nanometers. A safety condition is added to the tip vertical $z$-position controller to prevent the crashing into the hBN flake beyond the graphene edge : if the $z$-position reaches a threshold (typically $3\:$nm below the $z$-position of the tip estimated close to the edge), the tip is withdrawn and the CITS ends. Imaging of the Kekul\'e-bond order was carried out in STM constant-height mode after tuning the graphene to charge neutrality with the back gate, at a bias voltage corresponding to the energy of the LL$_{0^+}$ peak (see Ref.~\cite{Coissard22} for details).

\subsection*{Theoretical simulations}

To compute the local density of states shown in Fig.~\ref{Fig3} we use the simulation software Kwant~\cite{groth_kwant_2014}.
First we create a honeycomb lattice in a square system of size $L_x \times L_y=130\times 130$, in units of graphene's lattice constant $a$.
The unit-cell for the Kekul\'{e} order is tripled compared to pristine graphene, and is defined by the reciprocal vectors $\mathbf{a}_1 =a(3\sqrt{3}/2, 3/2)$ and $\mathbf{a}_2 = a(-3\sqrt{3}/2, 3/2)$, see Fig.~\ref{Fig3}A. 
To calculate the local density of states $\rho(E,x)$ at a given energy $E$ and Kekul\'{e}  unit cell $x$, we average over the six sites weighted by the corresponding wave-function, $\rho(E,x) = \sum_{\alpha} |\psi_{\alpha}(x)|^2 \delta(E-E_{\alpha})$, where $\alpha$ runs over the six unit cell sites. 
We compute the local density of states spectra, shown in Fig.~\ref{Fig3}B as a colormap,  using the kernel polynomial method \cite{Weisse:2006go} with a target energy resolution of $\Delta E/t = 0.005$, and a magnetic field of $\phi/\phi_0 = 0.005$ in units of the magnetic flux $\phi_0 = h/e$.
The dashed line spectra of Fig.~\ref{Fig3}B maps are obtained for a finite nano-ribbon of width $L_y$, with an arm-chair edge parallel to the $\hat{y}$ direction, as in Ref.~\cite{Pyatkovskiy14}.
We allow the edge to be mis-aligned with the Kekul\'{e} lattice vectors, as observed experimentally in Fig.~\ref{Fig1}F and G.
Lastly, the solid lines in Fig.~\ref{Fig3}C show cuts of the local density of states spectra show in Fig.~\ref{Fig3}B.
The dashed lines are calculated adding a single disorder realization obtained by adding a random on-site potential $V_{\mathrm{dis}}$ at each site to the clean local density of states spectra described above. 
The disorder strength at each site is drawn from a uniform distribution in the interval $[-W,W]$ with $W/t=0.3$.

\section*{Data availability}
All data needed to evaluate the conclusions in the paper are present in the paper and/or the Supplementary Materials.

\section*{Acknowledgments}
We thank C. D\'{e}prez, B. Halperin, M. Feigelman, M. Goerbig, M. Guerra, D. Perconte, H. Vignaud, and W. Yang for valuable discussions. 
We thank F. Blondelle, D. Dufeu, Ph. Gandit, D. Grand, G. Kapoujyan, D. Lepoittevin, J.-F. Motte and P. Plaindoux for technical support in setting up of the experimental system. Samples were prepared at the Nanofab facility of the N\'eel Institute. 
This work has received funding from the European Union's Horizon 2020 research and innovation program ERC grants \textit{QUEST} No. 637815 and \textit{SUPERGRAPH} No. 866365, and the Marie Sklodowska-Curie grant QUESTech No. 766025. A.~G.~G. acknowledges financial support by the ANR under the grant ANR-18-CE30-0001-01~(TOPODRIVE).

\section*{Author contributions} 
AC fabricated the sample and performed the measurements. AC, AGG, CR, HS and BS analysed the data. AGG and CR conducted the theoretical analysis. LV assembled the STM microscope. KW and TT supplied the hBN crystals. FG provided technical support on the experiment. BS conceived and supervised the project, designed the experimental set-up, and wrote the paper with inputs from all co-authors.

\section*{Competing Interests} The authors declare that they have no competing financial interests.

\section*{References}

\bibliography{QHedge-BIB}

\clearpage


\clearpage
\onecolumngrid
\setcounter{figure}{0}
\setcounter{section}{0}
\renewcommand{\thefigure}{S\arabic{figure}}

\newpage

~\vspace{5em}
\part*{ \centering Supplementary Information}
\bigskip
\vspace{5em}

\section{Sample details and AFM mapping}
\label{secSamples}

\begin{figure*}[b!]
	\centering
	\includegraphics[width=1\linewidth]{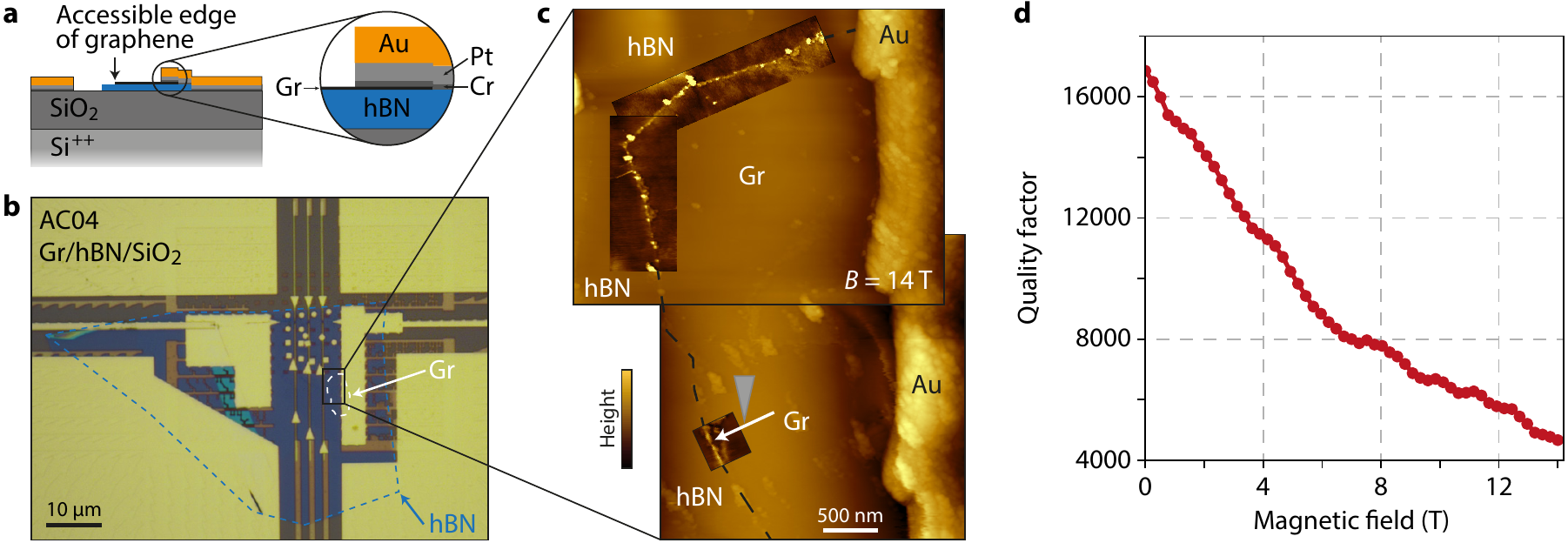}
	\caption{\textbf{Sample AC04 and AFM mapping.} \textbf{a}, The graphene/hBN heterostructure is deposited on a Si$^{++}$/SiO$_2$ substrate that serves as a back-gate electrode to tune the charge carrier density. Graphene is biased with a Cr/Pt/Au contact patterned by e-beam lithography on one of its edge, leaving others accessible for the tunneling spectroscopy of QH edge states. \textbf{b}, Optical image of the device. The graphene and hBN flakes are outlined by white and blue dashed lines, respectively. \textbf{c}, AFM mapping of the graphene sheet at $T=4.2\:$K and $B=14\:$T, with three high-resolution images of the edges. The gold contact used to bias graphene is visible on the right of the images. The edge studied in this work is indicated by the white arrow. \textbf{d}, Evolution of the quality factor $Q$ of the tuning fork with magnetic field.}
	\label{figS1}
\end{figure*}

The sample AC04 studied is this work is a heterostructure made of a graphene sheet atop a hexagonal boron nitride (hBN) flake, assembled by van der Waals stacking, and then deposited on a p$^{++}$Si/SiO$_2$ substrate to enable back gating of the charge carrier density in graphene. The voltage bias $V_\text{b}$ is applied using a Cr/Pt/Au contact patterned by e-beam lithography and covering partially the graphene sheet, leaving a large fraction of the perimeter accessible by the tip for imaging and tunneling spectroscopy of the edge states, see Fig. \ref{figS1}a and b. The graphene bulk properties of this sample have been presented in Ref.~\cite{Coissard22}.

The STM tip is brought atop the graphene sheet by AFM imaging of the coding markerfield patterned on the whole chip surface. This guiding process is done after about ten AFM images. An AFM mapping of graphene and its boundary with the underlying hBN performed at $B=14\:$T is shown in Fig. \ref{figS1}c. High-resolution AFM images of some edges are placed in overlay. These images reveal that the vacuum annealing employed to clean the graphene left some resist residues that have migrated toward the edges, forming bright spots in-between which edges are clean. In this work we focus on the edge indicated by the white arrow, which is also the direction of the Current Imaging Tunneling Spectroscopy (CITS) measurement grids performed from the bulk of graphene to the edge. Note that the tuning fork we used here still displays a relatively high quality factor in magnetic field, with $Q\sim 4000$ at 14~T (Fig. \ref{figS1}d).


\section{Localization of graphene edges on \lowercase{h}BN}

\begin{figure*}[ht!]
	\centering
	\includegraphics[scale=0.9]{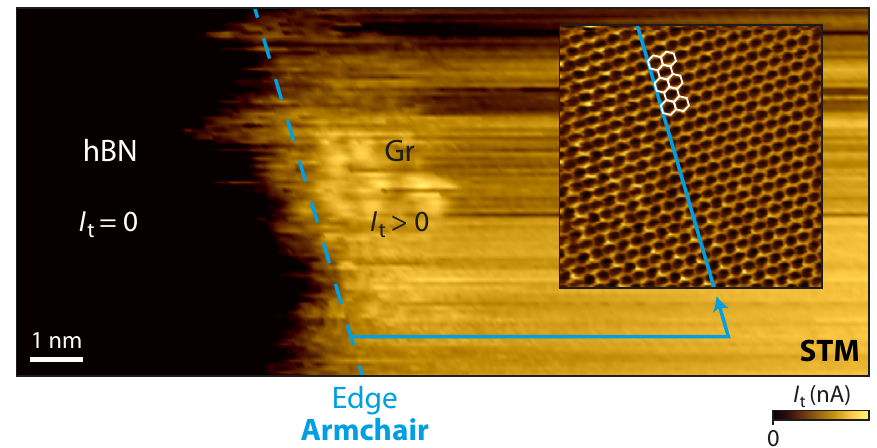}
	\caption{\textbf{Locating graphene armchair edge.} $16\times 7\:$nm$^2$ STM image in constant height mode of a graphene edge. On the left side the tunneling current vanishes, indicating the tip is atop hBN. Inset : $5\times 5\:$nm$^2$ STM image of the honeycomb lattice a few nanometers away from the edge. When reporting the edge direction (blue dashed line) on the honeycomb lattice, it coincides with an armchair edge orientation.}
	\label{figS2}
\end{figure*}

We show in Fig. \ref{figS2} a STM image at $B=14\:$T of the edge of graphene indicated by the white arrow in Fig. \ref{figS1}c, which provides a very accurate identification of the edge position. It is obtained in constant height mode: before STM imaging, we approach the STM tip in tunneling contact with the graphene in order to measure the setpoint tunneling current (typically $1\:$nA) and next switch off the $Z$-regulation for imaging. This mode allows a safe imaging of graphene edge since the tip would not crash down on the insulating hBN, but would rather simply measure zero tunneling current as seen on the left part of the STM image in Fig. \ref{figS2}. However, on the very edge of the graphene flake, the honeycomb lattice is not resolved due to the instability of the tunneling current at this location. As a result, the meaningful information in Fig. \ref{figS2} is the vanishing of the tunneling current when the tip reaches the hBN, which constitutes a clear identification of the edge location with nanometer-scale precision. Using this image of the edge, we can estimate its direction as indicated by the dashed blue line in Fig. \ref{figS2}. We report this line on the honeycomb lattice of the inset taken a few nanometers away and identify the armchair orientation for this edge.

We believe that the instability of the tunneling current measured on the very edge of the graphene in Fig. \ref{figS2} stems from the local lifting of the graphene sheet edge from the hBN flake, each time the STM tip scans over it, due to electrostatic interactions with the tip.


\section{Tip-induced lifting of the graphene edge and definition of the edge position}

We discuss here another way to locate the edge by means of a CITS grid spectroscopy measurement of the spatial dispersion of the Landau level (LL) spectrum toward the boundary. The grid spectroscopy is set to start far away in graphene bulk and to finish a few nanometers beyond the edge, previously located with STM images. Moreover, the slow $x$-axis direction of the grid is chosen to be perpendicular to the edge. A safety condition is added to the $Z$-controller to prevent the tip from crashing into hBN : if the $Z$-position of the tip goes below a threshold (typically $3\:$nm below the $Z$-position of the tip estimated close to the edge), the tip is withdrawn and the CITS ends.

\begin{figure}[t!]
	\centering
		\includegraphics[scale=1]{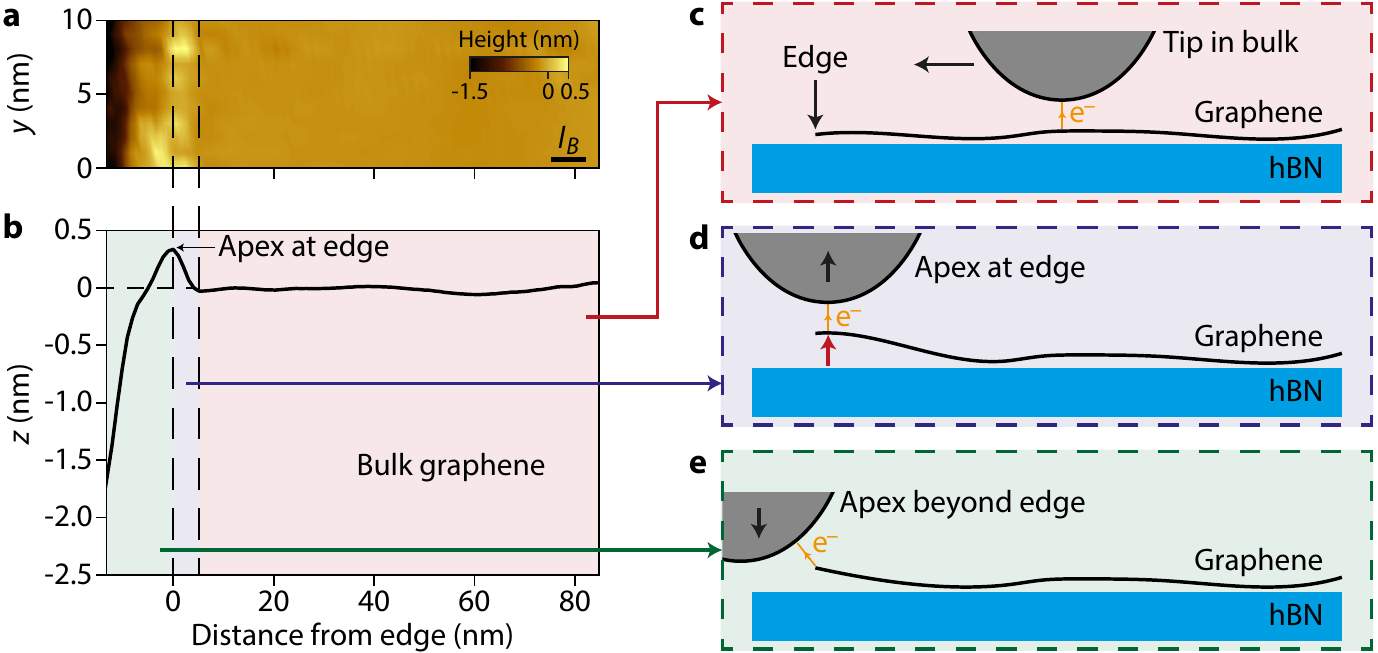}
	\caption{\textbf{CITS to the graphene edge: locating the edge.} \textbf{a,b}, $z(d_\text{edge},y)$ topographic map (a) and $z(d_\text{edge})$ profile (b) obtained from a CITS toward the armchair edge and beyond. We distinguish three regimes : the flat horizontal profile (red area) where the electrons tunnel in bulk graphene, see \textbf{c}, the sharp increase of $z$ (blue region) where the electrons tunnel from the tip apex very close to the edge in a situation where the sheet is strongly lifted, see \textbf{d}, and the decrease of $z$ when the tip apex is moved outside the graphene sheet, with a residual tunneling between the edge of graphene and other atoms on the side of the tip, see \textbf{e}, preventing the tip from crashing into hBN. The edge position is taken at the position of the maximum of the topographic profile.}
	\label{figS3}
\end{figure}

We show in Fig. \ref{figS3}a the topographic map $z(d_\text{edge},y)$ obtained from a CITS toward the graphene armchair edge identified in Fig. \ref{figS2}. $d_\text{edge}$ is the distance from the armchair edge, while $y$ is the lateral coordinate parallel to the edge. The topographic map features a clean and flat bulk graphene on a $80\times 10\:$nm$^2$ area next to the edge. When the tip is situated a few nanometers away from the edge, the $z(d_\text{edge},y)$ map reveals inhomogeneous bright spots. Though one can first think about residues, the small height of these spots, around $1-3\:\angstrom$, rules out this hypothesis.

We rather attribute these large spots to the lifting of the edge of the graphene sheet, as illustrated in Fig. \ref{figS3}d. The attractive van der Waals force of the tip was shown \cite{Georgi2017} to lift locally a graphene sheet lying on a SiO$_2$ substrate on a typical height of $1\:\angstrom$. Although we do not observe such lifting in Fig. \ref{figS3}a in bulk graphene (either because the deformation follows the tip such that we eventually observe an overall flat background, or because the deformation of the graphene sheet on hBN is more difficult, since the adhesion interactions between both materials are more important than between graphene and SiO$_2$), we can assume that the graphene flake  is more easily deformed at the edge by the force of the tip, and therefore the lifting is larger there than in the bulk.

The lifting of the edge is well visible in the height profile of Fig. \ref{figS3}b, obtained by averaging the topographic map along the $y$ direction (parallel to the edge). The $z$ profile features a flat region corresponding to bulk graphene (with variations of less than $1\:\angstrom$), and a hump of $3\:\angstrom$ height at the edge. After that, the tip quickly moves down by several nanometers until it meets the safety condition of the $Z$-controller, which stops the CITS. We attribute this lowering of the $z$ position to the fact that the tip apex has gone beyond the edge of graphene, but tunneling remains possible with some other higher atoms of the tip close to the apex, see Fig. \ref{figS3}d. This makes the measurement of a tunneling current possible even when the apex itself is lying on hBN, yet this current is highly unstable.

From this model we assume the position of the edge of graphene (\textit{i.e.} the tip apex is atop the edge) is given by the maximum of the hump in the $z$ profile, and from this origin we compute $d_\text{edge}$ the distance from the edge, which we use in the main text and the following figures.


\section{Additional tunneling conductance maps at the edge}

We show in this section two additional tunneling conductance maps acquired along the same armchair edge, but a few tens of nanometers away from the map shown in Fig. 2 of the main text. The back-gate voltage is fixed at $V_\text{g}=-5\:$V, corresponding to filling factor $\nu=0$.

\begin{figure}[p]
	\centering
		\includegraphics[width=0.6\textwidth]{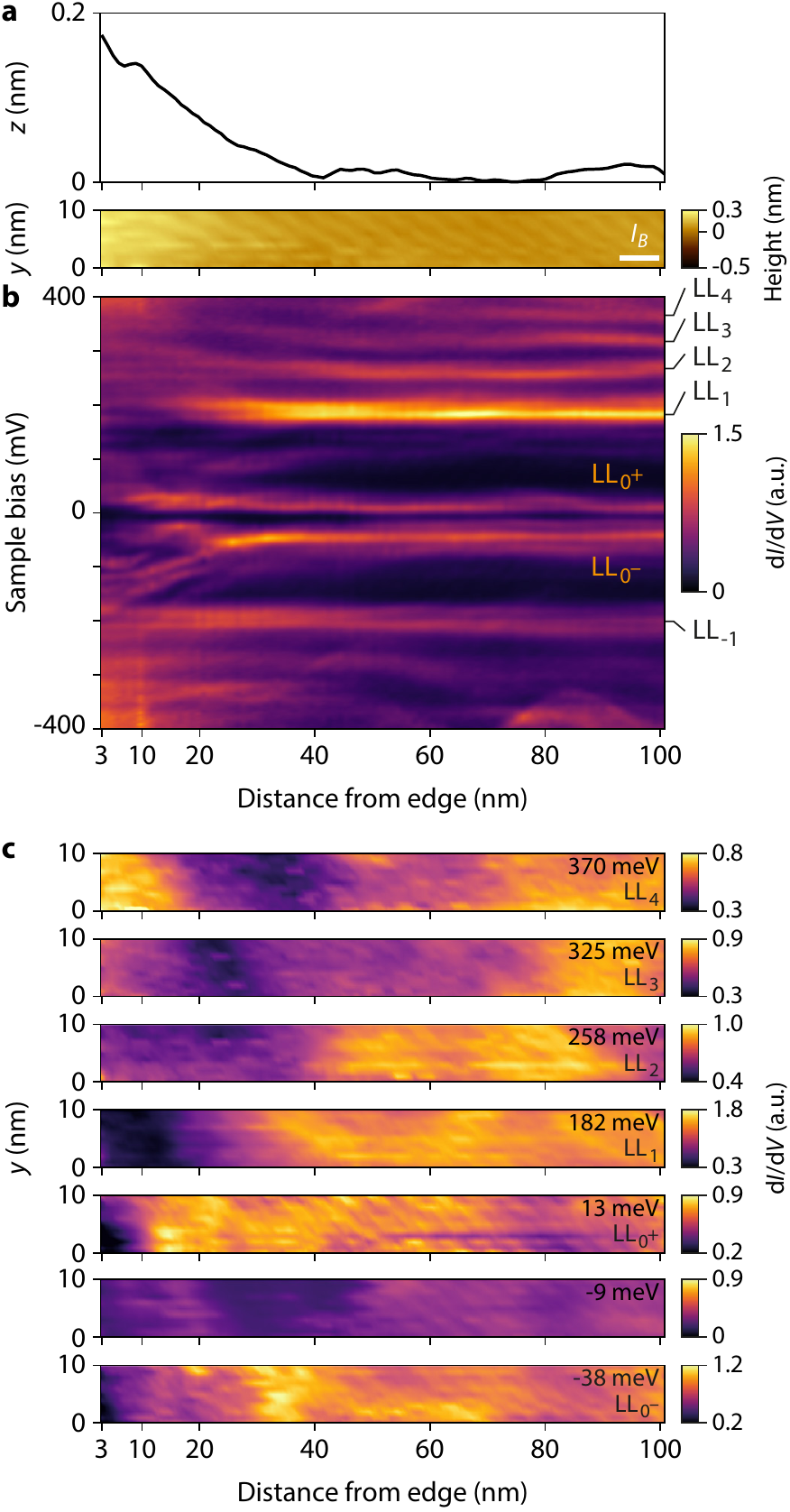}
	\caption{\textbf{Landau level spectroscopy toward armchair edge (1).} \textbf{a}, Topographic profile $z(d_\text{edge})$ and map $z(d_\text{edge},y)$ near the graphene edge ($l_B=6.85\:$nm at $B=14\:$T). \textbf{b}, Tunneling conductance as a function of the distance to the edge and sample bias, toward the edge. \textbf{c}, Tunneling conductance spatial maps at the different LL energies. The panel at $V_\text{b}=-9\:$mV shows the spatial map of the $\nu=0$ gap.}
	\label{figS4}
\end{figure}

\begin{figure}[ht!]
	\centering
		\includegraphics[width=0.75\textwidth]{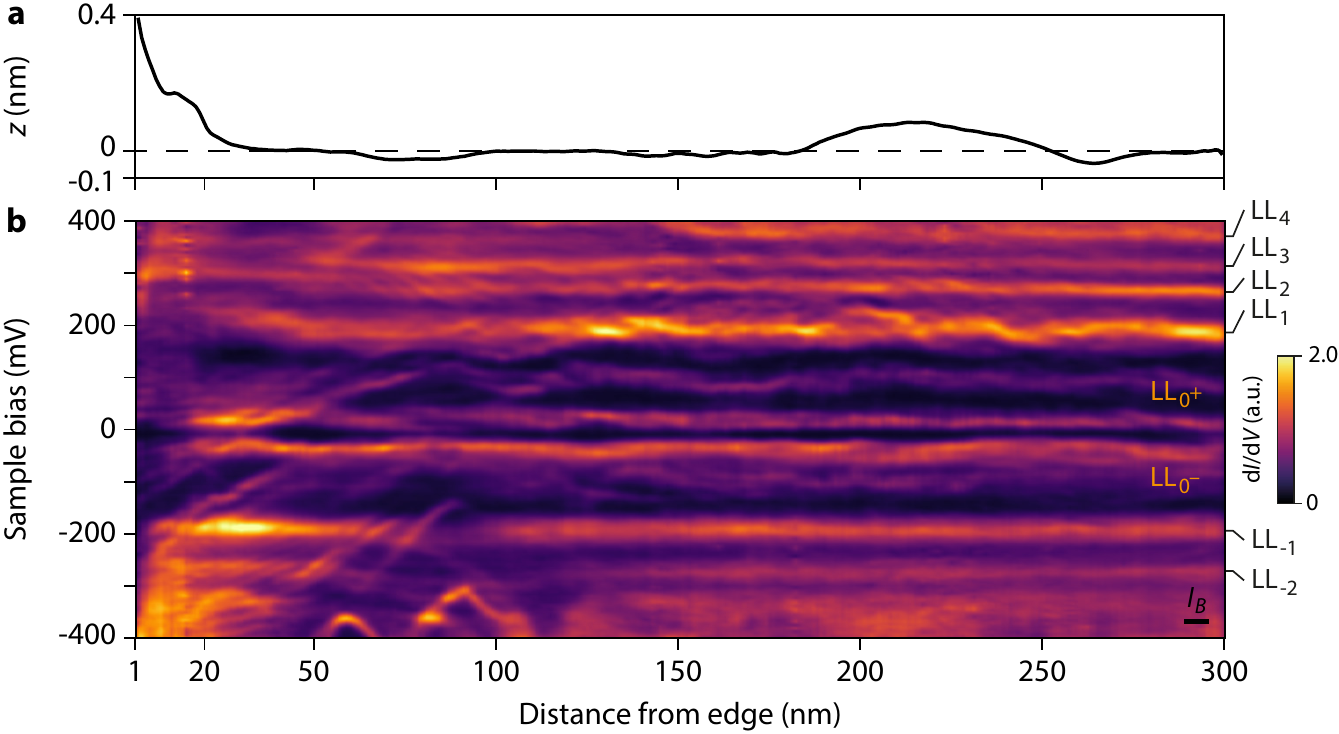}
	\caption{\textbf{Landau level spectroscopy toward armchair edge (2).} \textbf{a}, Topographic profile $z(d_\text{edge})$ near the graphene edge ($l_B=6.85\:$nm at $B=14\:$T). \textbf{b}, Tunneling conductance as a function of the distance to the edge and voltage bias.}
	\label{figS6}
\end{figure}

In Figs. \ref{figS4} and \ref{figS6}, the panels (a) show the topographic map $z(d_\text{edge},y)$ and the profile $z(d_\text{edge})$ obtained by averaging the map on the lateral $y$-dimension. Bulk graphene appears flat and clean, with a corrugation of at most  $1\:\angstrom$ on a distance of 100 and $300\:$nm, respectively. When approaching the edge on the left, $z(d_\text{edge})$ increases by around $2-3\:\angstrom$ due to the tip-induced lifting of the graphene sheet edge. In Fig. \ref{figS4} the CITS grid spectroscopy did not go beyond the edge: the edge position is rather roughly estimated using the STM image in Fig. 1d from the main text. The same goes for Fig. \ref{figS6}.

Panels (b) show the tunneling conductance toward the armchair edge as a function of the distance to the edge and the sample bias. The same qualitative observations as that of the main text can be made for the two edges: the Landau level peaks do not disperse when approaching the edges but vanish. The splitting of the LL$_0$ is well visible and the gap stays open down to the edge where it even gets more pronounced.

Panel (c) in Fig. \ref{figS4} shows the tunneling conductance as a function of the distance to the edge and the $y$-direction parallel to the edge, at different bulk Landau level energies $E_N$.

\begin{figure}[ht]
	\centering
		\includegraphics[width=0.65\textwidth]{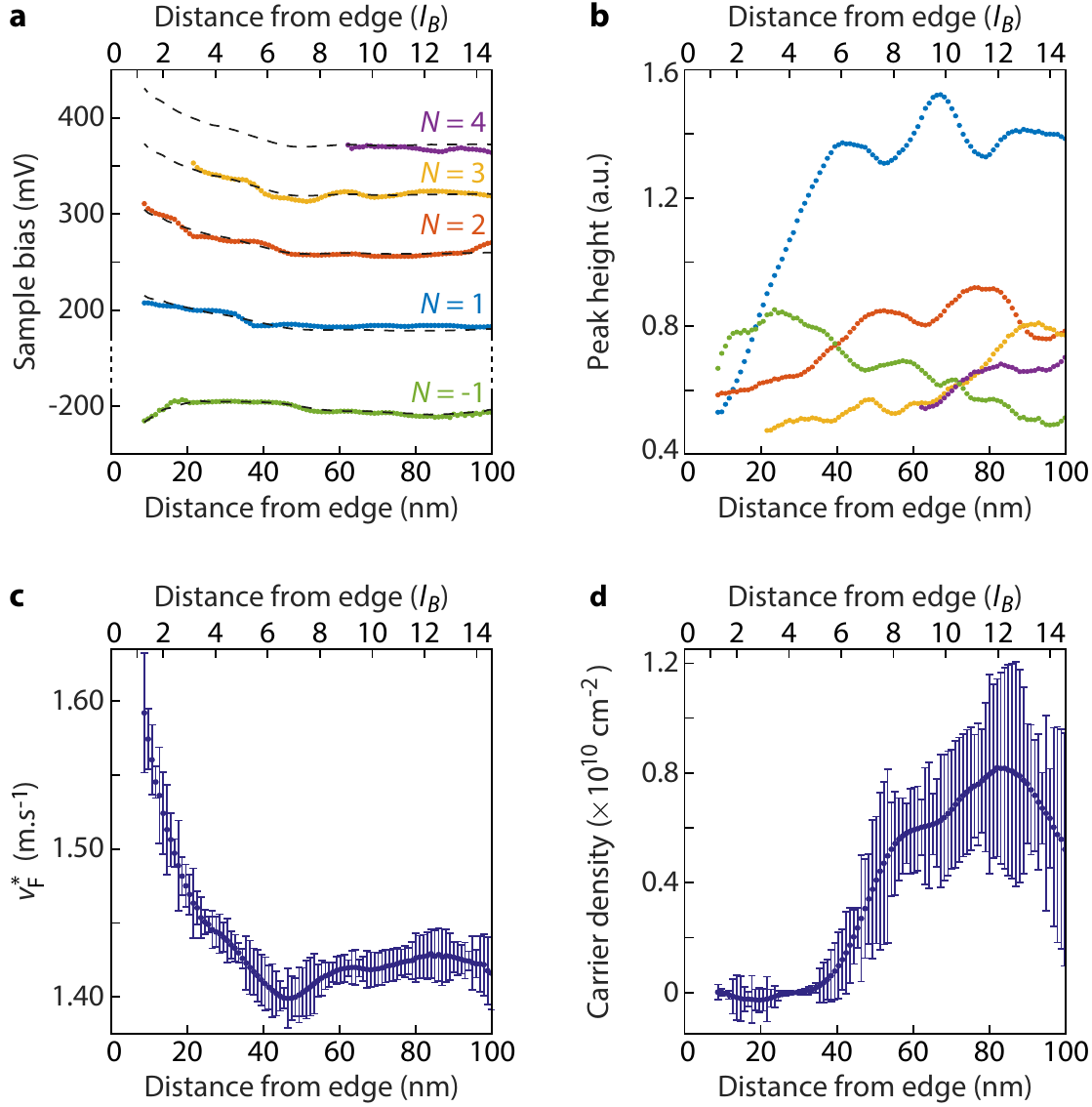}
	\caption{\textbf{a}, Peak energy for LL$_N$ as a function of the distance from the edge, extracted from the LDOS map in Fig. \ref{figS4}. The dashed lines show the fitted energies using the parameter $\vf^*$ obtained in (c). \textbf{b}, Peak height for the same LL$_N$ (same color code). \textbf{c}, Evolution of the effective Fermi velocity $\vf^*$ toward the edge obtained from the fit of LL$_{-1}$, LL$_{1}$, LL$_{2}$, LL$_{3}$ and LL$_{4}$ positions (when visible). \textbf{d}, Carrier density $n$ computed from the Dirac point position $E_\text{D}$ obtained from the same fit. $n$ presents a residual bulk value of $n_0\approx 7\times 10^{9}\:$cm$^{-2}$ and vanishes from $40\:$nm from the edge.}
	\label{figS7}
\end{figure}

We now consider in more details the tunneling conductance map shown in Fig. \ref{figS4}b. We plot in Fig. \ref{figS7}a the evolution of the positions in energy $E_N$ of the visible LL$_N$ peaks and in Fig. \ref{figS7}b the variation of their height as a function of $d_\text{edge}$. The amplitude of the peaks decreases as we approach the edge until peaks merge into a V-shape background at the edge where they are no longer visible. In particular, LL$_4$ vanishes at $9\,l_B$ from the edge, LL$_3$ at $3\,l_B$ whereas LL$_2$ and LL$_{\pm 1}$ disappear at $l_B$. The amplitude of LL$_1$ also vanishes way faster than the other LL$_N$ of higher index $N$. In addition to the peak vanishing at the edge, we can also notice in Fig. \ref{figS4}b and \ref{figS7}a a weak dispersion toward higher energy of the LL$_N$ peaks close to the edge (on a length of around $\sim 6\,l_B$ from the edge), see Ref.~\cite{Li2013}. 

Furthermore, we can fit the positions $E_N$ of LL$_{N\neq 0}$ at each $d_\text{edge}$ (for every visible LL at this point) with respect to equation $E_N=E_\text{D}+\text{sign}(N)\vf^{*}\sqrt{2\hbar e|N|B}$ to extract an effective Fermi velocity $\vf^*$ and an estimate of the Dirac point position $E_\text{D}$ as a function of $d_\text{edge}$. These results are shown in Figs. \ref{figS7}c for $\vf^*$ and \ref{figS7}d for $E_\text{D}$, which is converted into charge carrier density $n$ using $n=-\text{sign}(E_\text{D})\frac{1}{\pi}[E_\text{D}/\hbar\vf^*]^2$. The bulk value $v^*_\text{F,bulk}=1.42\times 10^6\:$m.s$^{-1}$ is consistent with a renormalization of the Fermi velocity due to the enhancement of electron-electron interactions at charge neutrality~\cite{DasSarma2007,Luican2011,Chae2012}, as characterized in a previous work \cite{Coissard22} for the same sample AC04. Below $7\,l_B$ the effective Fermi velocity starts to increase toward the armchair edge due to the dispersion of the LL peaks, reaching $v^*_\text{F,edge}=1.6\times 10^6\:$m.s$^{-1}$ at $l_B$ from the edge. As for the carrier density, we obtain a residual value $n_0\approx 7\times 10^{9}\:$cm$^{-2}$ in bulk graphene (in agreement with a back-gate voltage tuned at $\nu=0$). Below $60\:$nm the density is seen to decrease and eventually vanishes at $40\:\text{nm}=5\,l_B$ from the edge. A similar decrease of the density with respect to its bulk value has also been observed around $l_B$ from graphene edge on graphite~\cite{Li2013}. Finally, we use the $\vf^*(d_\text{edge})$ and $E_\text{D}(d_\text{edge})$ parameters to plot in Fig. \ref{figS7}a the fitted energies of each LL (black dashed lines). We notice a good agreement with the experimental points, especially for the dispersing parts.


\section{Tunneling conductance gate maps at different distances from the edge}

\begin{figure}[ht]
	\centering
		\includegraphics[width=0.85\textwidth]{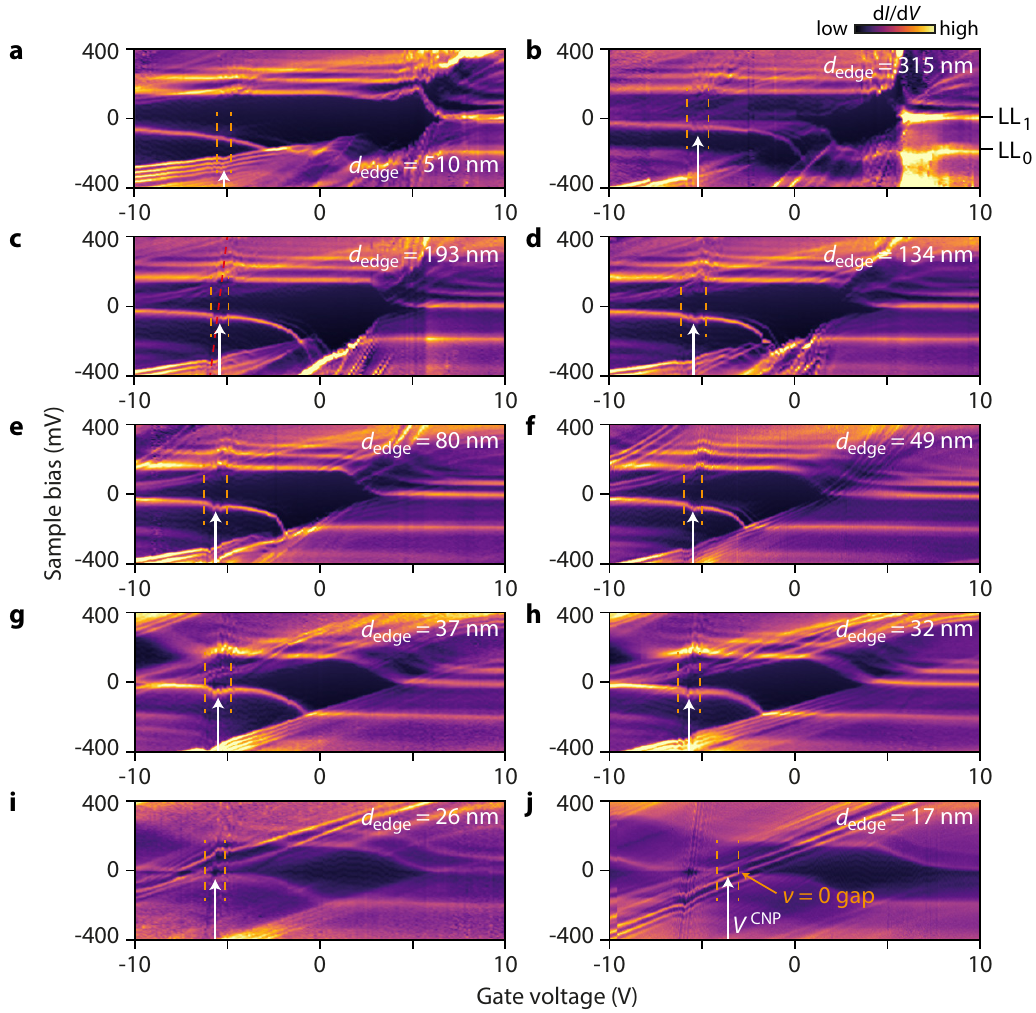}
	\caption{\textbf{Tunneling conductance gate maps at decreasing distances from the edge.} For every panel, we identify the $\nu=0$ gap that opens in charge-neutral graphene as a kink in the LL$_0$ peak when it pins the Fermi level at zero sample bias. The range of gate voltage where the gap opens is indicated by yellow dashed lines. The charge-neutrality point $V_\text{g}^\text{CNP}$ is then assumed to be in the middle of this range, where the $\nu=0$ gap is maximal, see white arrow. In panels (a-h), only the LL$_{0-}$ peak is well visible, the LL$_{0^+}$ peak is hindered. In panel (a), the kink at charge-neutrality is hardly visible for LL$_0$, we rather identify it in the charging peaks below it.}
	\label{figS8}
\end{figure}

We show in this section additional tunneling conductance gate maps (Fig. \ref{figS8}) used to plot the evolution of the charge-neutrality point $V_\text{g}^\text{CNP}$ as a function of the distance from the edge in Fig. 3c of the main text. $V_\text{g}^\text{CNP}$ is estimated at the middle of the $\nu=0$ gap opening when LL$_0$ pins the Fermi level at zero bias. This gap due to exchange interaction is indeed expected to be maximal at charge neutrality (i.e. half filling).

In each panel we indicate the opening of the $\nu=0$ gap by yellow dashed lines, observed as :
\begin{itemize}
\item[-] either as a typical gap opening between both LL$_{0^\pm}$ such as in panel (i,j),
\item[-] either as a kink toward negative energies in the LL$_{0^-}$ peak around zero sample bias, when the LL$_{0^+}$ is not visible, such as in panels (b-h),
\item[-] either as a kink in the other LL peaks or charging peaks when not easily visible for LL$_0$, such as in panel (a).\vspace{-1em}
\end{itemize}

Note that the opening of the $\nu=0$ gap at zero sample bias also induces a shift in energy of other LL peaks (and also of charging peaks), which enables unambiguous identification of the charge-neutrality point. Still these shifts in energy do not occur strictly at constant gate voltage due to tip-induced gating, see for instance the red dashed line in Fig. \ref{figS8}c.

\end{document}